# A New Key Establishment Scheme for Wireless Sensor Networks

Eric Ke Wang, Lucas C.K.Hui and S.M.Yiu

Department of Computer Science, the University of Hong Kong, Hong Kong
{kwang2,hui,smyiu}@cs.hku.hk

*ABSTRACT*

*Traditional key management techniques, such as public key cryptography or key distribution center (e.g., Kerberos), are often not effective for wireless sensor networks for the serious limitations in terms of computational power, energy supply, network bandwidth. In order to balance the security and efficiency, we propose a new scheme by employing LU Composition techniques for mutual authenticated pairwise key establishment and integrating LU Matrix with Elliptic Curve Diffie-Hellman for anonymous pathkey establishment. At the meantime, it is able to achieve efficient group key agreement and management. Analysis shows that the new scheme has better performance and provides authenticity and anonymity for sensor to establish multiple kinds of keys, compared with previous related works.*

*KEYWORDS*

*LU matrix composition, Key management, Key establishment, Wireless sensor networks*

## 1. INTRODUCTION

With the rapid development and wide application of wireless sensor networks (WSN), more and more security problems are emerging. Especially for the sensor networks deployed in a hostile environment with various potential malicious attacks, the security issue is a top priority concern. To ensure security in a wireless sensor network, it is essential to encrypt messages and authenticate the communicating nodes. Therefore, it is a major concern how to bootstrap secure communications between sensor nodes, i.e., how to set up secret keys between communicating nodes.

Since the key establishment is the initial step for secure communication, topics on key establishment, agreement and management has been studied for a long time. However, most of traditional key management schemes can not be applied to the WSN for the communication and computational limitations [1]. Currently, three main problems of WSN security are becoming hot. The first one is that how to enforce efficient security techniques for the resources-limited WSN. The second one is that how to possess the anonymity when the message are passed on multi-hop media. And the third one is that how to establish secure group communication inside the WSN [2].

### 1.1. Resources Constraints

Because of size, form factor and cost considerations, wireless sensor networks suffer from severe resource constraints, such as communication bandwidth and range, computation power, memory and energy. Therefore, WSN has a demand of the energy-efficiency of key establishment protocols. Traditional key establishment mainly includes public key cryptography which requires heavy computations. New techniques should be developed for the special conditions of WSN[3].

### 1.2. Secure Path Key Agreement

Considering that the sensor nodes are dispersed when they are deployed, nodes sometimes have to communicate others by the routing on multi-hops of ad hoc network. That means if two un-neighbored nodes are to agree on secret key with each other (We call it "path key agreement"), they have to send their secret respective message to each other over the multi-hop media. Especially if the network is deployed in a hostile environment, the process of path key agreement should not expose any sensitive information to the other parties even in an supposed secure channel.

### 1.3. Secure Group Communication

To conserve power, intermediate network nodes should aggregate results from individual sensors. Aggregation collects results from several sensors and calculates a smaller message that summarizes the important information from a group of sensors. For example, suppose the operator is interested in the average sensor reading for some value in the network. An inefficient way to find this would be for every sensor node to send its reading to the base station (possibly over multiple forwarding hops), and for the base station to calculate the average of all readings received. A more efficient way to collect the same information would be for intermediate nodes to forward the calculated average value of the readings they receive along with a count of the number of readings it incorporates. Each node then calculates the average for all of its descendents and only need send that value and the number of descendants to its parent. We call it "in-network process"[4][5]. This group operation needs secure group communications. However, group key establishment is the bottle neck for secure group communication in WSN.

The roadmap of this paper is as follows: Section 2 gives a brief overview of research background related to our scheme, including LU composition for pairwise key establishment,integrating LU composition with elliptic curve D-H for path-key establishment,and tree-based extension of LU composition for group key establishment . Section 3 illustrates the related works about this key establishment for WSN. Section 4 describes our scheme to solve the problem. Lastly, Section 5 gives some concluding remarks about this paper, as well as briefly reporting the status of this piece of research work.

## 2 BACKGROUND

### 2.1. LU Matrix Key

Definition 1. If the multiplication result of a n*n lower triangular matrix L and a n*n upper triangular matrix U equals to a symmetric matrix K, namely K=LU. We say K is the "LU composition" of triangular matrices L and U.

$$\begin{bmatrix} K_{11} & K_{12} & K_{13} \\ K_{21} & K_{22} & K_{23} \\ K_{31} & K_{32} & K_{33} \end{bmatrix} = \begin{bmatrix} A_{11} & 0 & 0 \\ A_{21} & A_{22} & 0 \\ A_{31} & A_{32} & A_{33} \end{bmatrix} \times \begin{bmatrix} B_{11} & B_{12} & B_{13} \\ 0 & B_{22} & B_{23} \\ 0 & 0 & B_{33} \end{bmatrix}$$

Figure 1. Key Matrix

### 2.2. Elliptic Curve Diffie-Hellman

Elliptic Curve Diffie-Hellman (ECDH) [7] is a key agreement protocol that allows two parties to establish a shared secret key over an insecure channel. Suppose Alice wants to establish a shared key with Party Bob, but the only channel available for them may be eaversdropped by a third party. Both have agreed to a common and publicly known curve $K$ over a finite field eg. $F_2^p$ as well as to a base point P €K in advance.

1. Alice randomly chooses $k_A$, $1 < k < 2^p$ and Bob accordingly $k_B$, $1 < k < 2^p$. Now $k_A$ is considered as Alice's private key, $k_B$ is Bob's private key.
2. Alice computes her public key: $Q_A = k_A P$, Bob does: $Q_B = k_B P$.
3. Alice sends $Q_A$ to Bob, Bob sends $Q_B$ to Alice.
4. Alice can now compute the shared secret for her and Bob by equation secret = $k_A Q_B$ and Bob also by secret = $k_B Q_A$. An eavesdropper knows only $Q_A$ and $Q_B$ but is not able to compute the secret from that.

The details about the algorithm principle of ECDH can be referenced in [27].

### 2.3. Tree-based Group Key Agreement for Ad-hoc Networks

A common model for WSN is that sensors (or nodes) in the network are deployed in clusters [8]. We follow this model and assume that sensors in the same partition are close to each other and more likely to be neighbours. The most popular ways to agree on keys and solve the continues key management problem is to build a binary tree structure in each node [9]. Therefore, it can achieve efficient key management.

## 3. RELATED WORKS

According to the secure communication requirement in WSN, two kinds of key establishment are needed. One is pairwise key establishment, the other is group key establishment. A few schemes has been proposed which consist of three phases in general [10]:(1)*key setup* prior to deployment, (2) *shared-key discovery* after deployment, and (3) *path-key establishment* if two sensor nodes do not share a direct key.

The most popular pairwise key pre-distribution solution is *Random Pairwise Key Scheme* [11] which addresses unnecessary storage problem and provides some key resilience. It is based on Erdos and Renyi's [12] work. Each sensor node stores a random set of $Np$ pair-wise keys to achieve probability $p$ that two nodes are connected. Neighboring nodes can tell if they share a common pair-wise key after they send and receive "Key Discovering" Message within radio range. Its defect is that it sacrifices key connectivity to decrease the storage usage. *Closest (location-based) pair-wise keys pre-distribution scheme* [13] is an alternative to Random pair-wise key scheme. It takes advantage of the location information to improve the key connectivity. Later on, *Random key-chain based key pre-distribution solution* is another random key pre-distribution solutions which originated from the solution of basic probabilistic key pre-distribution scheme [14]. It relies on probabilistic key sharing among the nodes of a random graph.

There are several key reinforcement proposals to strengthen security of the established link keys, and improve resilience. Objective is to securely generate a unique link or path key by using established keys, so that the key is not com- promised when one or more sensor node is captured. One approach is to increase amount of key overlap required in shared key discovery phase. Q-composite random key pre-distribution scheme [11] requires q common keys to establish a link key. Similar mechanism is proposed by Pair-wise key establishment protocol [15] which uses threshold secret sharing for key reinforcement. The key reinforcement solutions in general increase processing and communication complexity, but provide good resilience in the sense that a compromised key-chain does not directly affect security of any links in the WSN. But, it may be possible for an adversary to re- cover initial link keys. An adversary can then recover reinforced link keys from the recorded multi-path reinforcement messages when the link keys are compromised.

Actually, due to the randomness of the key selection process in Random key pre-distribution schemes, none of the above key management schemes can guarantee that a pairwise key will be found between two nodes wanting to communicate, and when the number of compromised nodes exceeds a certain threshold their security decreases drastically. To address this issue, key matrix based dynamic key generation solutions introduced. All possible link keys in a network of size N can be represented as an N N key matrix. It is possible to store small amount of information to each sensor node, so that every pair of nodes can calculate corresponding field of the matrix, and uses it as the link key. The original idea is from Blom's scheme [16]. Based on this work, [6] proposed a LU matrix key pre-distribution for WSN. It can guarantee that any pair of nodes can find a pairwise key between themselves. This is achieved by using the secret information assigned from the lower and upper triangular matrixes decomposed from a symmetric matrix of a pool of keys. Because all the established pairwise keys are distinct to each other, any sensor's compromise cannot affect the secure communication between non-compromised nodes. However, it cannot guarantee anonymity when two nodes exceed radio range want to establish the path-key.

Considering in-network process such as data aggregation, we need to explore the way to build a secure group communication for WSN. Several group key management protocols have been proposed for mobile ad-hoc group communication. The protocol of [17] provides efficient mutual authentication and group key agreement for low-power mobile devices, and supports dynamic changes. However, it requires a wireless infrastructure with some powerful trusted server (base station) that performs heavy computations, which is not allowed according to the trust model in WSN. [18], [19], [20] proposed some key management and related schemes for wireless ad hoc network. Unfortunately, they all assume that after sensors are deployed, they are considered static. Thus it cannot solve the problem of key management when new members join or old members leave. To solve this problem, tree-based schemes are introduced to make key management more effective and efficient. [21] has analyzed and optimized a number of CGKA protocols (i.e., BD [22], CLIQUES [23], STR [24] ,TGDH [25] and [29, 30,31,32]) from the perspective of static and dynamic mobile ad-hoc groups. The optimized way relies on the the tree-based extension of the well-known elliptic curve Diffie-Hellman key exchange protocol (ECDH). However, public key cryptography solutions are still not the first choice for WSN since its high computational overhead.

## 4.TKLU scheme(tree-based key exchange protocol with LU Matrix Composition)

### 4.1 Overview

Tree-based key exchange protocol with LU Matrix Composition(TKLU) has three protocols which are Pairwise key establishment protocol, Path Key establishment protocol and group key establishment protocol. (1)Sensor nodes who are neighboring can establish pairwise key after they are deployed by LU composition technique. And they are able to authenticate each other in the process of pairwise key establishment. (2)Sensor nodes who are not neighboring should establish secret keys over the multi-hop path. They can achieve to agree on keys in unsecure channel. Even if the third parties obtain the message they can not deduce the keys. It also can achieve authenticity at the same time. (3)Those neighborhood nodes can agree on group key for secure data aggregation.

Table 1.  Notations.

| A,B,C,….M1,…Mn. | Node identities |
|---|---|
| $SM$ | symmetric matrix |
| GF(q) | finite filed |
| $Aix$ | The row of node i |

| *Ayi* | The colum of node i |
|---|---|
| *Kij* | Mediate value of keys |
| *Tii* | Groups |
| $r_i$ | Random number |
| *KCi* | A key chain maintained by node i |

## 4.2 Phase 1:Key Pre-distribution

*Step 1: Base Generation*

1. The base station generates a large pool of keys(e.g. $5^{20}$ or more). The keys are selected from a finite filed GF(q) to create a symmetric matrix(*SM*). Where *q* is the smallest prime larger than the key size.
2. The base station select one publicly known curve *K* over a finite field eg. $F_2^p$ as well as to a base point *P∈K*.

*Step2:Decompose Matrices to obtain LU Matrices*

Base station does the decomposition of the created *SM* to obtain one lower triangular matrix *A* and one upper triangular matrix *B*.

*Step3: Key Pre-distribution*

Every node is randomly assigned one row from matrix *A* and one corresponding column from matrix *B*. For example, node *i* is assigned row *Aix* and column *Byi*, node *j* is assigned row *Ajx* and column *Byj*. After the key pre-distribution, each node only has two vectors in its memory. Each vector has *n* elements.

## 4.3 PHASE 2: KEY ESTABLISHMENT

*Pairwise Key Establishment Protocol*

After key pre-distribution, each node can establish a pairwise key with its neighbors to make sure the secure around communication.
We design a protocol for the process of pairwise key establishment:
   1. Node *i* sends its column *Byi* to node j.
   2. After node *j* receive *Byi*, it computes *Kji* by vector multiplication of *Bjx* and *Byj* .
   3. Node *j* reply the *Byj* ,*F(Kji)* where *F(Kji)* is the Hash result of the computation of the last step.
   4.Upon receiving *Byj* , node *i* compute *Kij* and check if *F(Kij) = F(Kji)*.
   5.If it is verified, node *i* send *F(Kij)* to node *j* for the verification.

*Path Key Establishment Protocol*

Suppose node *i* and node *j* are not neighboring to each other. If they need to establish a secure communication channel, the first thing they have to do is to establish a path key between them. The steps are as follows( Figure 2):

1. Node *i* generates a random number $r_i$, where $1 < r_i < 2^p$, compute $Q_i = r_i P$. And node *i* sends *Byi;Qi* to node *j* over routing.

2. Upon node *j* receive the path key request from node *i*, node *j* generates a random number *rj* , where $1 < r_j < 2^p$, compute $Q_j = r_{ij}P$. At the same time, node *j* computes *Kji* by vector multiplication of *B*$_{yi}$ and *B*$_{yj}$ .

3. After that, node *j* reply the message *B*$_{yj}$ ; *F(Kji)*;*Qj* to node *i*. If receiving *B*$_{yj}$ ;*Qj* , node *i* compute *Kij* and check if *F(Kij)* = *F(Kji)*. If it is verified, it computes *Qij* =*riQj* and reply a message with *F(Kij)*; *F(Qij)*. Otherwise, it broadcast an error message.
4. Upon receiving *F(Kij)*; *F(Qij)*, node *j* check if *F(Kji)* = *F(Kij) and F(Qji) = F(Qij)*.

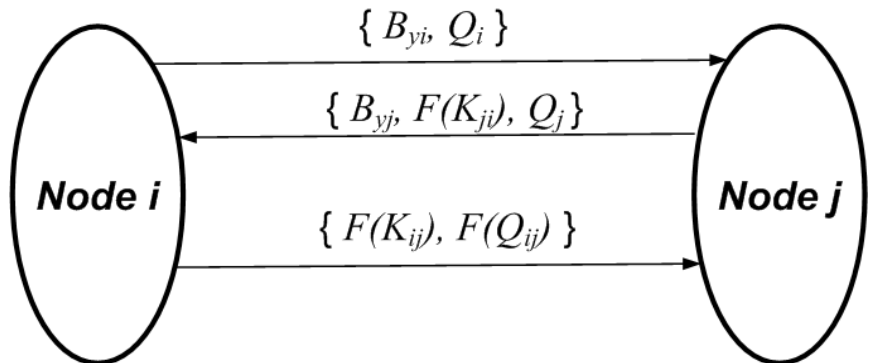

Figure 2 Path Key Establishment Protocol

Through the protocol, node *i* and *j* can establish a path key *Qij* (*or Qji*)with authenticating each other. The biggest merit is that any other parties cannot deduce the final path key even they can eavesdrop the channel for the discrete logarithm problems.

*Group Key Establishment Protocol*

As figure 3 shows, suppose there are *n* members who wish to form a group (key tree). The shallowest rightmost leaf node is chosen as the sponsor. First, all members are ordered according to some criteria to be $M_1$, $M_2$... ,$M_n$. Then the first two members, $M_1$ and $M_2$, execute a 2-party pairwise key exchange presented above. The same do the members $M_i$ and $M_{i+1}$, where *i* is an even integer from [0, n]. Now we have *n*=2 groups. Each sponsor broadcasts then its tree with all blinded keys. The rest process is similar to be the one in the first round, if we consider a member as a group with only one member. Such process is repeated until all members are in one group. After *ith* round the number of groups is reduced to *n=2i*. The setup of the group is finished after *logn*2 rounds. An example with n= 6 as shown in figure 3. Assume the order of members is $M_1$, $M_6$. In the first round $M_1$ and $M_2$ perform a pairwise key establishment and forms a new group $T_{I1}$ . Similarly, groups $T_{I2}$ of $M_3$ and $M4$, and $T_{I3}$ of $M_5$ and $M_6$ are formed respectively. In the second round, $M_2$, $M_4$ broadcasts their trees with respective blinded keys. Then groups $T_{I1}$ and $T_{I2}$ form a new group $T_{II1}$, $T_{I3}$ do nothing and is renamed to $T_{II2}$ . Finally the only two groups $T_{II1}$ and $T_{II2}$ form the group $T_{III}$ consisting of all members. Then a group key tree is built up.

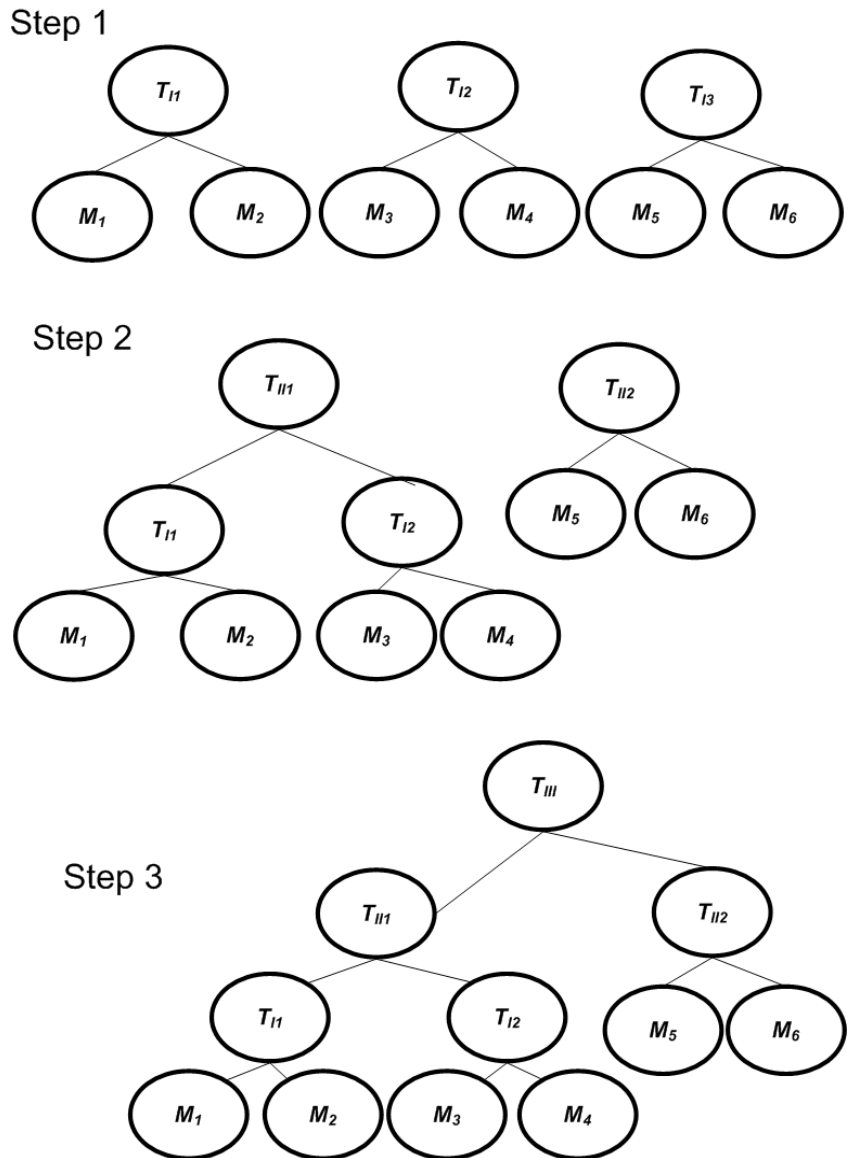

Figure 3 Group Key Establishment Protocol

## 5. DISCUSSION AND PERFORMANCE ANALYSIS

### 5.1 Discussion for Security

Compromised Keys Revocation Compromised keys should be removed from the wireless sensor networks immediately. In our proposed scheme, each node knows its neighbor's ID and shares a unique pairwise communication key only with an intended 1-hop neighbor. Those nodes between several-hop distance shares path keys. Each node also shares some group keys inside its group after the nodes are deployed. Once a misbehaving node which was captured or compromised by the adversary is detected (we do not discuss malicious node detection in this paper. Reader can reference [26] or other related articles), all its 1-hop neighbor nodes immediately remove the corresponding pairwise keys shared with it. At the same time, the leader in the group including the malicious node broadcasts the group key revocation process to update group keys. Other nodes which have shared path keys with it remove those path keys instantly. Moreover, the misbehavior node ID will be sent to the sink node to broadcast the entire network about this node removal message. Each node needs to check whether it has keys shared with this node.
Another security concern is resiliency to Node Capture. Once adversaries capture sensor nodes, they can obtain the secret information inside such as communication keys and other sensitive data. Actually, node capture attach is the biggest threat in wireless sensor networks. On the area of key establishment for wireless sensor networks, several key pre-distribution schemes are very popular. The main idea is that each sensor node has m keys which are randomly drawn from a key pool. If two nodes have one common key, they will create a communication link by using the key. The most popular random key pre-distribution schemes are from Eschenauer's random key pre-distribution scheme [14] and Adrian's "q-composite" scheme [11];Comparing with those schemes, we can achieve zero fraction of compromised keys in non compromised sensor nodes with the increasing numbers of sensors.

### 5.2 MEMORY ANALYSES

The tradition solutions for key establishment has two ways. One extreme solution in terms of resource usage is to deploy single master key to all sensors. Since, an adversary may capture a node and compromise the key very easily, it has very low resilience. The other extreme is to use distinct pair-wise keys for all possible pairs in the WSN. For a network of size $N$, each sensor *node i* ($1<=i<=N$) should pre-stores a key-chain $KC_i = \{K_{ij}|i \neq j$ and $1<=j<=N\}$ of size $N$-1 out of $N(N-1)=2$ distinct keys. Node *i* stores a unique pair-wise key for each one of $N$-1 sensor nodes in the WSN. Actually, this solution is a very exhaustive one which creates unnecessary storage burden on a sensor node although this solution has very good key resilience.

Our scheme can perform better memory storage than pure pairwise key pre-distribution schemes. Before the deployment, sensor nodes are preloaded with 2 vectors which have $N$ figures(While, in most of other pairwise key establishment schemes, keys are preloaded into sensor nodes). After deployment, for dense sensor fields (suppose average $k$ neighbors per node), one node needs stores $k$ pairwise keys. In addition, Suppose naturally the deployment of sensors divides the sensors into $m$ groups, on average, each group has $n = N/m$ ($N$ is the network size) sensors inside. Thus each sensor should have $k$ pairwise keys and Log(n) group keys(here path key can be ignored as a small number). The number of keys in each sensor
should be $k + Log(N/m)$. For a very high-density networks such as dense sensor fields (about 40 neighbors per node), a high-density group which consists of 50 to 80 nodes(suppose the whole number of a general network sensors is 2000), one node only needs store around 40 + 7 = 47 keys after key establishment.

### 5.3 EXPERIMENT

We implement our scheme on MICA2. The Berkeley MICA2 mote has a 7.3MHz processor with 128 KB flash memory, 4.0 KB RAM, and a Chipcon CC1000 radio at 19.2 Kbps. Our experiments employ Tiny OS (TOS) as the sensor operating system [6, 4], executing on MICA2 motes. We tested the key setup time for the two key setup protocols: pairwise key establishment, path key establishment and group key establishment. Here, we used two to twenty motes to execute our key setup schemes. Every node can communicate with every other node. This simulates the different densities of sensor networks, from each node having only one neighbor to each node having nine neighbors.

Figure 4 shows the total time for our key establishment schemes. First, we see that as the network becomes denser, the time for pairwise key establishment grows longer. For example, for a two node network, the time for pairwise key setup is about 0.21 seconds, and for a ten node network, the time is about 2.77 seconds. Another finding is that the number of messages for the key setup scheme will significantly affect the completion time for that protocol as the density increases.

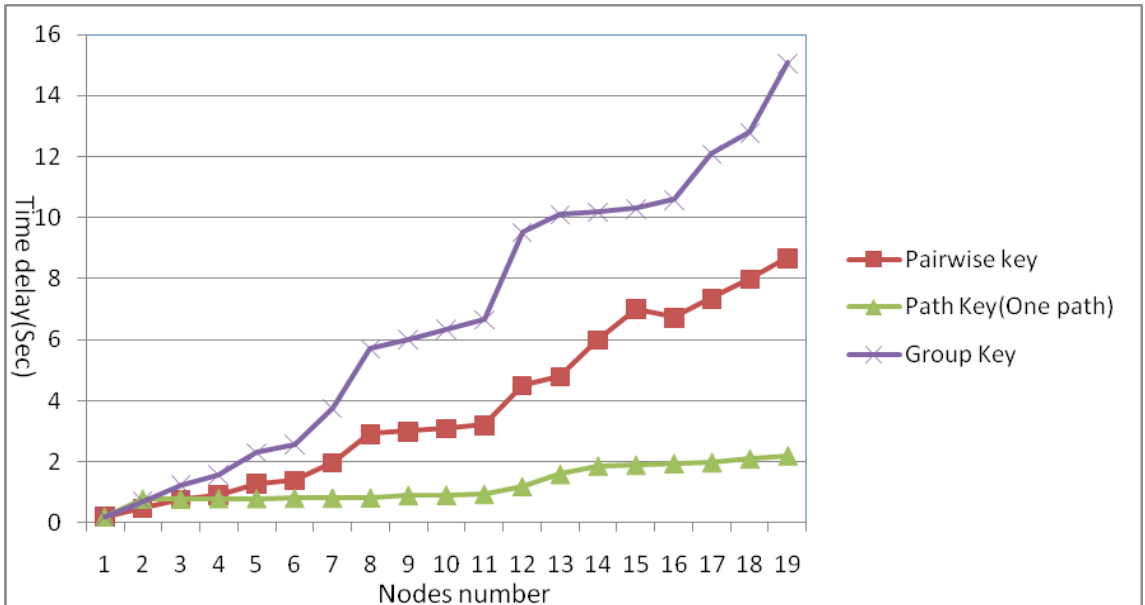
Figure 4  Time delay of Key establishment

To provide more details, Table 2 shows the total time delay of seting up pairwise keys and group keys, and average time delay of path keys.  This table confirms two observations from figure 4: 1) Path keys and pairwise keys establishment are relative low cost; 2) It still cost less than the new key management schemes [28, 29, 30].

Table 2.  Time delay of key establishment (Second).

| Nodes number | Pairwise key (total time) | Path key (Average) | Group key (Total time) |
|---|---|---|---|
| 2 | 0.21 | 0.21 | 0.21 |
| 3 | 0.48 | 0.79 | 0.71 |
| 4 | 0.78 | 0.79 | 1.25 |
| 5 | 0.93 | 0.81 | 1.58 |
| 6 | 1.301 | 0.81 | 2.31 |
| 7 | 1.402 | 0.82 | 2.58 |
| 8 | 1.98 | 0.825 | 3.77 |
| 9 | 2.92 | 0.83 | 5.71 |
| 10 | 3.01 | 0.91 | 6.02 |
| 11 | 3.11 | 0.92 | 6.35 |
| 12 | 3.21 | 0.95 | 6.67 |
| 13 | 4.52 | 1.2 | 9.53 |
| 14 | 4.81 | 1.6 | 10.11 |
| 15 | 6.01 | 1.87 | 10.2 |
| 16 | 7.01 | 1.89 | 10.3 |
| 17 | 6.73 | 1.95 | 10.6 |
| 18 | 7.05 | 1.98 | 12.1 |
| 19 | 7.33 | 2.1 | 12.8 |
| 20 | 7.9 | 2.2 | 15.08 |

## 5. CONCLUSION AND FUTURE WORK

In this paper, we introduce a new scheme that can be used for establish various keys(pairwise keys, path keys and group keys) for wireless sensor networks. It can achieve quick authenticity without extra computations and communications. The experiment result shows the performance of TKLU is invigorating.

**Authors**

Eric Ke Wang is a Ph.D. candidate in computer science department from the University of Hong Kong. His research interests mainly involve information security, data mining, Network Security, E-commerce. He has been a research assistant in French National Institute for Research in Computer Science and Control( (INRIA),Nancy，France, in 2007.

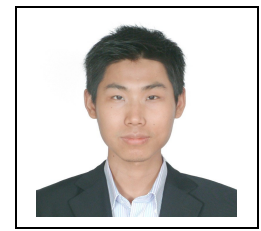